\documentclass[runningheads]{llncs}

\begin{document}

\title{Sorting a Low-Entropy Sequence}
\author{Travis Gagie}
\institute{Department of Computer Science\\
	University of Toronto\\
	\email{travis@cs.toronto.edu}}
\maketitle

\vspace{-20ex}
\begin{center}
STUDENT PAPER
\end{center}
\vspace{17ex}

\begin{abstract}
We give the first sorting algorithm with bounds in terms of higher-order
entropies: let $S$ be a sequence of length $m$ containing $n$ distinct
elements and let \(H_\ell (S)\) be the $\ell$th-order empirical entropy of 
$S$, with \(n^{\ell + 1} \log n \in O (m)\); our algorithm sorts $S$ 
using \((H_\ell (S) + O (1)) m\) comparisons.
\end{abstract}

\section{Introduction}
\label{introduction}

Sorting in the comparison model is one of oldest problems in computer
science, but it remains an important and active area.  Previous research
has shown how we can take advantage of various kinds of pre-sortedness,
such as long runs, few inversions, or only a small number of elements out
of place (see~\cite{PM95}); in this paper, we show how we can take
advantage of low entropy to reduce comparisons.

Consider a fixed sequence \(S = s_1, \ldots, s_m\) containing $n$ distinct
elements drawn from a total order.  For any non-negative integer $\ell$,
the \emph{$\ell$th-order empirical entropy} of $S$, denoted \(H_\ell
(S)\), is our expected uncertainty about $s_i$ (measured in bits) given a
context of length $\ell$, as in the following experiment: we are given
$S$; $i$ is chosen uniformly at random from \(\{1, \ldots, m\}\); if \(i
\leq \ell\), we are told $s_i$; if \(i > \ell\), we are told \(s_{i -
\ell}, \ldots, s_{i - 1}\).  Specifically,
\[H_\ell (S)
= \left\{ \begin{array}{ll}
\displaystyle \sum_{a \in S} \frac{\#_a (S)}{m} \log \frac{m}{\#_a (S)}
\hspace{3ex} & \mbox{if \(\ell = 0\);} \\
& \\
\displaystyle \frac{1}{m} \sum_{\alpha \in A_\ell} |S_\alpha| H_0 (S_\alpha)
\hspace{3ex} & \mbox{if \(\ell > 0\).}
\end{array} \right.\]
Here, \(a \in S\) means $a$ occurs in $S$; \(\#_a (S)\) is the number of
occurrences of $a$ in $S$; $\log$ means $\log_2$; $A_\ell$ is the set of
$\ell$-tuples in $S$; and $S_\alpha$ is the sequence whose $i$th element 
is the one immediately following the $i$th occurrence of $\alpha$ in $S$.  
The length of $S_\alpha$ is the number of occurrences of $\alpha$ in $S$
unless $\alpha$ is a suffix of $S$, in which case it is 1 less.

Notice \(\log n \geq H_0 (S) \geq \cdots \geq H_{m - 1} (S) = H_m (S) = 
\cdots = 0\).  For example, if $S$ is the string TORONTO, then \(\log n = 
2\),
\begin{eqnarray*}
H_0 (S)
& = & \frac{1}{7} \log 7
	+ \frac{3}{7} \log \frac{7}{3}
	+ \frac{1}{7} \log 7
	+ \frac{2}{7} \log \frac{7}{2}
\approx 1.84\ ,\\
H_1 (S)
& = & \frac{1}{7} \left( \rule{0ex}{2ex}
	H_0 (S_\mathrm{N})
	+ 2 H_0 (S_\mathrm{O})
	+ H_0 (S_\mathrm{R})
	+ 2 H_0 (S_\mathrm{T})
	\right) \\
& = & \frac{1}{7} \left( \rule{0ex}{2ex}
	H_0 (\mathrm{T})
	+ 2 H_0 (\mathrm{RN})
	+ H_0 (\mathrm{O})
	+ 2 H_0 (\mathrm{OO})
	\right) \\
& = & 2 / 7 \approx 0.29
\end{eqnarray*}
and all higher-order empirical entropies of $S$ are 0.  This means, if
someone chooses a character uniformly at random from TORONTO and asks us
to guess it, then our uncertainty is about \(1.84\) bits.  If they tell us
the preceding character before we guess, then on average our uncertainty
is about \(0.29\) bits; if they tell us the preceding two or more
characters, then we are certain of the answer.  The difference between
0th-order and higher-order empirical entropies can be of practical
importance: the encodings produced by most older compression algorithms
are only bounded in terms of the 0th-order empirical entropy of the input,
whereas those produced by most modern compression algorithms are bounded
in terms of higher-order empirical entropies.  For example,
Manzini~\cite{Man01} proved Burrows and Wheeler's algorithm~\cite{BW94}
encodes $S$ using at most
\[(8 H_\ell + O (1)) m + N^\ell (2 N \log N + 9)\]
bits, where $\ell$ is any non-negative integer, $N$ is the size of the 
alphabet and, depending on the implementation, the hidden constant is 
about \(2 / 25\).

Suppose we want to sort $S$, that is, to put the elements of $S$ in
non-decreasing order.  Many familiar sorting algorithms already take
advantage of low 0th-order empirical entropy: Munro and Spira~\cite{MS76}
proved MergeSort, TreeSort and HeapSort use \((H_0 (S) + O (1)) m\)
ternary comparisons\footnote{A ternary comparison of $x$ and $y$ tells us
whether \(x < y\), \(x = y\) or \(x > y\); a binary comparison only tells
us whether \(x \leq y\) or \(x > y\).  Our algorithm uses binary
comparisons, which is a slight advantage: while most instruction sets
support ternary comparisons, most high-level languages do not; a ternary
comparison is usually implemented as two binary
comparisons~\cite{And91}.}; by the Static Optimality Theorem~\cite{ST85},
SplaySort uses \(O ((H_0 (S) + 1) m)\) comparisons; Sedgewick and
\mbox{Bentley}~\cite{SB02} recently proved QuickSort uses \(O ((H_0 (S) +
1) m)\) comparisons in the expected case.

In Section~\ref{simple_section} we give a new algorithm that sorts $S$
using \((H_0 (S) + O (1)) m\) comparisons.  In 
Section~\ref{parameterized_section} we generalize it so that, given a
non-negative integer $\ell$ with \(n^{\ell + 1} \log n \in O (m)\), it
uses \((H_\ell (S) + O (1)) m\) comparisons.  Our algorithm's main
disadvantage is its slowness: it takes \(O ((H_\ell (S) + 1) m \log n +
\ell m)\) time, whereas the algorithms mentioned above take \(O ((H_0 (S)
+ 1) m)\) time. It works in models where, for \(t \leq m\), it takes \(O
(\log t)\) time to perform a standard operation on a balanced binary
search tree with $t$ keys, each of \(O (\log m)\) bits~\cite{CLRS01}; if
such a tree takes \(O (t)\) space, then our algorithm takes \(O (m)\)  
space.  We emphasize that we do not make assumptions about the source of
$S$, nor do we use randomization or pointer arithmetic.

\section{Sorting $S$ using $(H_0 (S) + O (1)) m$ Comparisons}
\label{simple_section}

If we are given a list of the distinct elements in $S$ and their
frequencies, then we can easily sort $S$ using fewer than \((H_0 (S) + 2)
m\) comparisons: we construct a nearly optimal leaf-oriented binary search
tree $T$, as described in Subsection~\ref{mehlhorn_subsection}, and
perform an insertion sort into $T$.  A \emph{leaf-oriented} binary search
tree (LBST) is one in which the data are stored at the leaves.

Since we are not given that information, we instead start with an LBST
$T_1$ on $s_1$; for $i$ from 2 to $m$, we search for $s_i$ in $T_{i - 1}$
and then ``in effect'' construct a new LBST $T_i$ which is nearly optimal
for \(s_1, \ldots, s_i\).  In Subsection~\ref{sequence_subsection} we
prove this uses \((H_0 (S) + O (1)) m\) comparisons.  Of course, actually
constructing every $T_i$ would be very slow; in
Subsection~\ref{statistics_subsection}, we show how we can quickly ``in
effect'' construct them.  We used a similar approach in~\cite{Gag04} for 
dynamic alphabetic coding.

\subsection{Constructing a Nearly Optimal Leaf-Oriented Binary Search Tree}
\label{mehlhorn_subsection}

Let \(a_1, \ldots, a_n\) be the distinct elements in $S$ in increasing
order.  By Shannon's Noiseless Coding Theorem~\cite{Sha48}, if we search
for \(s_1, \ldots, s_m\) in an LBST on \(a_1, \ldots, a_n\), then we use
at least \(H_0 (S) m\) comparisons.  Mehlhorn~\cite{Meh77} gave an \(O
(n)\)-time algorithm that, given \(a_1, \ldots, a_n\) and \(\#_{a_1} (S),
\ldots, \#_{a_n} (S)\), constructs an LBST with which we use fewer than
\((H_0 (S) + 2) m\) comparisons; we follow Knuth's~\cite{Knu98}
presentation.

\begin{theorem}[Mehlhorn, 1977]
\label{mehlhorn_theorem}
We can construct a leaf-oriented binary search tree on \(a_1, \ldots, 
a_n\) whose leaves have depths \(\left\lceil \log \frac{m}{\#_{a_1} (S)} 
\right\rceil + 1, \ldots, \left\lceil \log \frac{m}{\#_{a_n} (S)} 
\right\rceil + 1\).
\end{theorem}

\begin{proof}
For \(1 \leq i \leq n\), let
\[f_i
= \sum_{j = 1}^{i - 1} \frac{\#_{a_j} (S)}{m} + \frac{\#_{a_i} (S)}{2 m}\ .\]
Since \(|f_i - f_{i'}| > \frac{\#_{a_i} (S)}{2 m}\) for \(i' \neq i\), the
first \(\left\lceil \log \frac{m}{\#_{a_i} (S)} \right\rceil + 1\) bits of
$f_i$'s binary representation suffice to distinguish it; let $\sigma_i$ be
this sequence of bits.  Notice \(\sigma_1, \ldots, \sigma_n\) are
lexicographically increasing.

We construct a binary tree such that, for \(1 \leq i \leq n\) and \(1 \leq
k \leq |\sigma_i|\), the $k$th edge on the path from the root to the $i$th
leaf is a left edge if the $k$th bit of $\sigma_i$ is a 0, and a right
edge if it is a 1.  We store \(a_1, \ldots, a_n\) at the leaves.  At each
internal node $v$, if $v$ has two children, then we store a pointer to the 
rightmost leaf in $v$'s left subtree.
\qed
\end{proof}

Consider the LBST this algorithm produces.  When searching for $s_i$, we
start at the root and descend to the leaf that stores $s_i$, as follows:
at each internal node $v$, if $v$ has two children and the rightmost leaf
in $v$'s left subtree stores element $a$, then we compare $s_i$ with $a$ 
and proceed to $v$'s left child or right child depending on whether \(s_i 
\leq a\); if $v$ has only one child, we proceed immediately to that child.  
Searching for \(s_1, \ldots, s_m\), we use a total of at most
\[\sum_{i = 1}^n \#_{a_i} (S)
	\left( \left\lceil \log \frac{m}{\#_{a_i} (S)} \right\rceil + 1 \right)
< (H_0 (S) + 2) m\]
comparisons.

\subsection{Using a Sequence of Leaf-Oriented Binary Search Trees}
\label{sequence_subsection}

Let $F$ be the set of indices $i$ such that $s_i$ is the first occurrence
of that element in $S$; that is, \(F = \{i\ :\ s_i \not \in s_1, \ldots,
s_{i - 1}\}\).  For \(1 \leq i \leq m\), let $T_i$ be the nearly optimal
LBST Mehlhorn's algorithm constructs for \(s_1, \ldots, s_i\), augmented
so that, for \(a \in s_1, \ldots, s_i\), the leaf storing $a$ also stores
a counter set to \(\#_a (s_1, \ldots, s_i)\) and a list containing the
indices of $a$'s occurrences in \(s_1, \ldots, s_i\).  Consider the
concatenation of the lists in $T_m$, as a permutation: its inverse sorts
$S$.\,\footnote{In fact, if each list in $T_m$ is in increasing order,
then their concatenation's inverse stably sorts $S$; a \emph{stable} sort
preserves equal elements' relative order.} Thus, with regard to
comparisons needed, constructing $T_m$ is equivalent to sorting $S$; the
following lemmas show \((H_0 (S) + O (1)) m\) comparisons suffice.

\begin{lemma}
\label{T_m_lemma}
We can construct $T_m$ using at most
\[\sum_{i \in F - \{1\}} (\lceil \log (i - 1) \rceil + 3) +
	\sum_{i \not \in F} \left( \left\lceil \log \frac{i - 1}
	{\#_{s_i} (s_1, \ldots, s_{i - 1})} \right\rceil + 3 \right)\]
comparisons.
\end{lemma}

\begin{proof}
By induction.  We can construct $T_1$ without using any comparisons.  For 
\(2 \leq i \leq m\), suppose we have $T_{i - 1}$ and want to construct 
$T_i$.  To do this, we first search for $s_i$ in $T_{i - 1}$.

If \(s_i \in s_1, \ldots, s_{i - 1}\), that is, \(i \not \in F\), then
our search uses \(\left\lceil \log \frac{i - 1}{\#_{s_i} (s_1, \ldots,
s_{i - 1})} \right\rceil + 1\) comparisons and ends at the leaf storing
$s_i$.  Otherwise, our search uses at most \(\lceil \log (i - 1) \rceil +
1\) comparisons and ends at a leaf storing either $s_i$'s predecessor or
successor in $T_{i - 1}$.

Let $a$ be the element stored at the leaf $v$ where our search ends.  We
determine whether $a$ is $s_i$'s predecessor, $s_i$ itself, or $s_i$'s
successor by checking whether \(a \leq s_i\) and whether \(s_i \leq a\).  
If $a$ is $s_i$'s predecessor, then we insert a new leaf immediately to
the right of $v$, that stores $s_i$, a counter set to 1 and a list
containing $i$; if \(a = s_i\), we increment $v$'s counter and add $i$ to
$v$'s list; if $a$ is $s_i$'s successor, then we insert a new leaf
immediately to the left of $v$, that stores $s_i$, a counter set to 1 and
a list containing $i$.

Notice $T_{i - 1}$'s leaves now contain the same information as $T_i$'s,
which is enough for us to construct $T_i$ without any further comparisons.  
In total, if \(i \not \in F\), then we use \(\left\lceil \log \frac{i -
1}{\#_{s_i} (s_1, \ldots, s_{i - 1})} \right\rceil + 3\) comparisons to
construct $T_i$; otherwise, we use at most \(\lceil \log (i - 1) \rceil +
3\) comparisons.
\qed
\end{proof}

\begin{lemma}
\label{analysis_lemma}
\[\sum_{i \in F - \{1\}} \log (i - 1) + \sum_{i \not \in F}
	\log \frac{i - 1}{\#_{s_i} (s_1, \ldots, s_{i - 1})}
\leq (H_0 (S) + O (1)) m\ .\]
\end{lemma}

\begin{proof}
Let
\begin{eqnarray*}
C & = & \sum_{i \in F - \{1\}} \log (i - 1) + \sum_{i \not \in F}
	\log \frac{i - 1}{\#_{s_i} (s_1, \ldots, s_{i - 1})} \\
& < & \log (m!) - \sum_{i \not \in F} \log \#_{s_i} (s_1, \ldots, s_{i - 1})\ .
\end{eqnarray*}
For \(i \not \in F\), if $s_i$ is the $j$th occurrence of $a$ in $S$, then 
\(j \geq 2\) and \linebreak \(\log \#_{s_i} (s_1, \ldots, s_{i - 1}) = 
\log (j - 1)\).  Thus,
\begin{eqnarray*}
C
& < & \log (m!) - \sum_{a \in S} \sum_{j = 2}^{\#_a (S)} \log (j - 1) \\
& = & \log (m!) - \sum_{a \in S} \log (\#_a (S)!) +
	\sum_{a \in S} \log \#_a (S) \\
& \leq & \log (m!) - \sum_{a \in S} \log (\#_a (S)!) + n \log \frac{m}{n} \\
& = & \log (m!) - \sum_{a \in S} \log (\#_a (S)!) + O (m)\ .
\end{eqnarray*}
By Stirling's Formula,
\[x \log x - x \ln 2
< \log (x!)
\leq x \log x - x \ln 2 + O (\log x)\ .\]
Thus,
\[C
\leq m \log m - m \ln 2 - \sum_{a \in S} \left( \rule{0ex}{2ex}
	\#_a (S) \log \#_a (S) - \#_a (S) \ln 2 \right) + O (m)\]
Since \(\sum_{a \in S} \#_a (S) = m\),
\hfill \raisebox{-12ex}[0ex][0ex]{\qed}
\begin{eqnarray*}
C & \leq & \sum_{a \in S} \#_a (S) \log \frac{m}{\#_a (S)} + O (m) \\
& = & (H_0 (S) + O (1)) m\ .
\end{eqnarray*}
\end{proof}

\subsection{Using a Statistics Data Structure}
\label{statistics_subsection}

Let \(T_1, \ldots, T_m\) be as defined in
Subsection~\ref{sequence_subsection}.  Since Mehlhorn's algorithm takes
\(O (n)\) time time, sorting $S$ by constructing \(T_1, \ldots, T_m\)
takes \(O (m n)\) time; this is faster than BubbleSort, for example, but
still impractical.  To save time, we implement all of the $T_i$s as a
single dynamic \emph{statistics} data structure: an augmented balanced
binary search tree that stores a list of triples \(\langle a_1, w_1, L_1
\rangle, \ldots, \langle a_t, w_t, L_t \rangle\), each of which consists
of a key $a_j$, a positive integer weight $w_j$ and a list $L_j$.  None of
the following operations compares keys and each takes \(O (\log t)\)  
time~\cite{CLRS01}:
\pagebreak
\begin{description}
\item[\(\mathbf{search} (b)\):]
	return the smallest $j$ with \(\sum_{k = 1}^j w_k \geq b\);
\item[\(\mathbf{sum} (j)\):]
	return \(\sum_{k = 1}^j w_k\);
\item[\(\mathbf{triple} (j)\):]
	return \(\langle a_j, w_j, L_j \rangle\);
\item[\(\mathbf{increment} (j)\):]
	increment $w_j$;
\item[\(\mathbf{append} (i, j)\):]
	append $i$ to $L_j$;
\item[\(\mathbf{insert} (a, i, j)\):]
	insert \(\langle a, 1, \langle i \rangle \rangle\)
	into the $j$th position in the list of triples.
\end{description}
As an aside, we note there are faster statistics data structures on a word
RAM (see~\cite{PD04}); we leave as future work investigating whether we
can improve our algorithm with one of them.

\begin{lemma}
\label{statistics_lemma}
Suppose we have a statistics data structure whose keys and weights are,
respectively, the distinct elements in \(s_1, \ldots, s_i\) and their 
frequencies.  Then given the path from the root to a node $v$ in $T_i$,
we can determine the following in \(O (\log n)\) time:
\begin{itemize}
\item if $v$ is a leaf, the element stored at $v$;
\item whether $v$ has a left child;
\item whether $v$ has a right child;
\item if $v$ has two children,
	the element stored at the rightmost leaf in $v$'s left subtree.
\end{itemize}
\end{lemma}

\begin{proof}
Let \(a_1, \ldots, a_t\) be the distinct elements in \(s_1, \ldots, s_i\) 
in increasing order and, for \(1 \leq j \leq t\), let
\[f_j
= \sum_{k = 1}^{j - 1} \frac{\#_{a_k} (s_1, \ldots, s_i)}{i}
	+ \frac{\#_{a_j} (s_1, \ldots, s_i)}{2 i}\ .\]
Given a binary string $\rho$, we can find the smallest $j$ such that
$f_j$'s binary representation begins $\rho$, if one exists: let $j'$ be
the value returned by \(\mathbf{search} ((.\rho) i)\) with $.\rho$
interpreted as a binary fraction; by $T_i$'s construction, we know the $j$
we seek is either $j'$ or \(j' + 1\); we use \(\mathbf{sum} (j')\),
\(\mathbf{triple} (j')\) and \(\mathbf{triple} (j' + 1)\) to compute
$f_{j'}$ and $f_{j' + 1}$, if they are defined.

Let $\sigma$ be the path from the root to $v$ encoded as a binary string,
with each 0 indicating a left edge and each 1 indicating a right edge.  We
can determine each of the following properties of $v$ in \(O (\log t)
\subseteq O (\log n)\) time: if there is only one $j$ such that $f_j$'s
binary representation begins $\sigma$, then $v$ is a leaf storing $a_j$;
if $v$ is an internal node and there is a $j$ such that $f_j$'s binary
representation begins \(\sigma 0\), then $v$ has a left child; similarly,
if $v$ is an internal node and there is a $j$ such that $f_j$'s binary
representation begins \(\sigma 1\), then $v$ has a right child; finally,
if $v$ has two children, then there is a $j$ such that $f_j$'s binary
representation begins \(\sigma 0\) and $f_{j + 1}$'s binary representation
begins \(\sigma 1\) --- the rightmost leaf in $v$'s left subtree stores 
$a_j$.
\qed
\end{proof}

For \(2 \leq i \leq m\), let \(a_1, \ldots, a_t\) be the distinct elements
in \(s_1, \ldots, s_{i - 1}\).  Suppose we have a statistics data
structure $D$ implementing $T_{i - 1}$, that is, storing \(\langle a_1,
\#_{a_1} (s_1, \ldots, s_{i - 1}), L_1 \rangle, \ldots, \langle a_t,
\#_{a_t} (s_1, \ldots, s_{i - 1}), L_t \rangle\) with each $L_j$
containing the indices of $a_j$'s occurrences in \(s_1, \ldots, s_{i -
1}\).  Using $D$ and Lemma~\ref{statistics_lemma}, if \(s_i \in s_1,
\ldots, s_{i - 1}\), then searching for $s_i$ in $T_{i - 1}$ takes
\(\left\lceil \log \frac{i - 1}{\#_{s_i} (s_1, \ldots, s_{i - 1})}
\right\rceil + 1\) comparisons and \(O \left( \left( \log \frac{i -
1}{\#_{s_i} (s_1, \ldots, s_{i - 1})} + 1 \right) \log n \right)\) time,
and returns $j$ such that \(a_j = s_i\); otherwise, searching for $s_i$
takes at most \(\lceil \log (i - 1) \rceil + 1\) comparisons and \(O
((\log (i - 1) + 1) \log n)\) time, and returns $j$ such that $a_j$ is
either $s_i$'s predecessor or successor in $T_{i - 1}$.  Determining
whether $a_j$ is $s_i$'s predecessor, $s_i$ itself, or $s_i$'s successor
takes two more comparisons and \(O (\log n)\) time.  If $a_j$ is $s_i$'s
predecessor, then we use \(O (\log n)\) time to insert \(\langle s_i, 1,
\langle i \rangle \rangle\) into the \((j + 1)\)st position in the list of
triples; if \(a_j = s_i\), then we use \(O (\log n)\) time to increment
the weight in the triple \(\langle a_j, \#_{a_j} (s_1, \ldots, s_{i - 1}),
L_j \rangle\) and append $i$ to $L_j$;  if $a_j$ is $s_i$'s successor,
then we use \(O (\log n)\) time to insert \(\langle s_i, 1, \langle i
\rangle \rangle\) into the $j$th position in the list of triples.  After
this, $D$ implements $T_i$.

We can construct a statistics data structure implementing $T_1$ in \(O
(1)\) time without using any comparisons; by Lemmas~\ref{T_m_lemma}
and~\ref{analysis_lemma}, we can use \((H_0 (S) + O (1)) m\) comparisons
and \(O ((H_0 (S) + 1) m \log n)\) time to construct a statistics data
structure implementing $T_m$; from this we can obtain the concatenation of
the lists in $T_m$, in \(O (m \log n)\) time.  Therefore, we can sort $S$
using \((H_0 (S) + O (1)) m\) comparisons and \(O ((H_0 (S) + 1)  m \log
n)\) time.

\section{Sorting $S$ using $(H_\ell (S) + O (1)) m$ Comparisons}
\label{parameterized_section}

To generalize our algorithm, given $S$ and $\ell$ with \(n^{\ell + 1} \log
n \in O (m)\), we work from left to right and maintain a set of statistics
data structures, one for each distinct $\ell$-tuple seen so far, and keep
track of them using two dictionaries.  In effect, we partition $S$, use
the statistics data structures to sort each of the parts, and then merge
them.  This uses a total of \((H_\ell (S) + O (1)) m\) comparisons and
\(O ((H_\ell (S) + 1) m \log n + \ell m)\) time.

\subsection{Using a Set of Statistics Data Structures}
\label{set_subsection}

As we work, we maintain a statistics data structure $D_{\alpha}$ for each
distinct $\ell$-tuple $\alpha$ that has occurred so far.  Assume we have a
black box $B$ that works as follows: for \(\ell + 1 \leq i \leq m\),
suppose we query $B$ immediately before we process $s_i$; if the
$\ell$-tuple \(s_{i - \ell}, \ldots, s_{i - 1}\) has occurred before, then
$B$ returns a pointer to $D_{s_{i - \ell + 1}, \ldots, s_{i - 1}}$;
otherwise, $B$ creates $D_{s_{i - \ell + 1}, \ldots, s_{i - 1}}$ and
returns a pointer to it; in both cases, querying $B$ costs \(O (\log n)\)
comparisons and \(O (\ell \log n)\) time.

We use $B$ to keep track of the statistics data structures, but we only
query it after seeing a new distinct \((\ell + 1)\)-tuple; this way, the
total cost of querying $B$ is \(O (n^{\ell + 1} \log n) \subseteq O (m)\)
comparisons and \(O (n^{\ell + 1} \ell \log n) \subseteq O (\ell m)\)  
time.  We augment the statistics data structures so that, instead of
storing just triples, they store quadruples: each $D_{b_1, \ldots,
b_\ell}$ stores a list of quadruples \(\langle a_1, w_1, L_1, p_1 \rangle,
\ldots,\) \(\langle a_t, w_t, L_t, p_t \rangle\), where $p_j$ is a pointer
to $D_{b_2, \ldots, b_\ell, a_j}$; as well, $D_{b_1, \ldots, b_\ell}$
stores the ranks of \(b_2, \ldots, b_\ell\), which we define and use
later.

\pagebreak

To process $S$, first, we query $B$ to obtain a pointer to a new
statistics data structure $D_{s_1, \ldots, s_\ell}$.  For \(\ell + 1 \leq
i \leq m\), we search for $s_i$ in the LBST that $D_{s_{i - \ell}, \ldots,
s_{i - 1}}$ implements, as in Subsection~\ref{statistics_subsection}.  If
\(s_{i - \ell}, \ldots, s_i\) has occurred before, then this search
returns a quadruple \(\langle s_i, w, L, p \rangle\); we increment $w$,
append $i$ to $L$, and retrieve $p$, which points to $D_{s_{i - \ell + 1},
\ldots, s_i}$.  If \(s_{i - \ell}, \ldots, s_i\) has not occurred before,
then we query $B$ to obtain a pointer $p$ to $D_{s_{i - \ell + 1}, \ldots,
s_i}$ and insert \(\langle s_i, 1, \langle i \rangle, p \rangle\) into
$D_{s_{i - \ell}, \ldots, s_{i - 1}}$.

As in Section~\ref{introduction}, let $A_\ell$ be the set of $\ell$-tuples 
in $S$ and, for \(\alpha \in A_\ell\), let $S_\alpha$ be the sequence 
whose $j$th element is the one immediately following the $j$th occurrence 
of $\alpha$ in $S$.  In total, processing $S$ takes
\[O (m) + \sum_{\alpha \in A_\ell}
	(H_0 (S_\alpha) + O (1)) |S_\alpha|
= (H_\ell (S) + O (1)) m\]
comparisons and
\[O (\ell m) + \sum_{\alpha \in A_\ell}
	O \left( \rule{0ex}{2ex} (H_0 (S_\alpha) + 1)
	|S_\alpha| \log n \right)
= O ((H_\ell (S) + 1) m \log n + \ell m)\]
time.  When we finish processing $S$, for each \(\alpha \in A_\ell\) and 
each \(a \in S_\alpha\), $D_\alpha$ contains a quadruple \(\langle a, \#_a
(S_\alpha), L, p \rangle\) with $L$ containing the indices of occurrences 
of $a$ that immediately follow occurrences of $\alpha$ in $S$.

After we process $S$, the at most $n^\ell$ statistics data structures each
contain at most $n$ quadruples.  Consider all these quadruples, as well as
``dummy'' quadruples \(\langle s_1, 1, \langle 1 \rangle, \mathrm{null}
\rangle, \ldots, \langle s_\ell, 1, \langle \ell \rangle, \mathrm{null}
\rangle\); we sort them all by their keys, which takes \(O ((n^{\ell + 1}
+ \ell) \log n) \subseteq O (m)\) comparisons and time.  Now consider the
concatenation of their lists of indices: its inverse sorts $S$.  To see
why, notice the indices are in non-decreasing order by the elements they
index.  Thus, to prove the following theorem, it only remains for us to 
implement our black box $B$.

\begin{theorem}
\label{sorting_theorem}
Given a sequence \(S = s_1, \ldots, s_m\) containing $n$ distinct elements 
and a non-negative integer $\ell$ with \(n^{\ell + 1} \log n \in O (m)\), 
we can sort $S$ using \((H_\ell (S) + O (1)) m\) comparisons and \(O 
((H_\ell (S) + 1) m \log n + \ell m)\) time.
\end{theorem}

\subsection{Using a Dictionary of Elements and a Dictionary of $\ell$-tuples}
\label{dictionary_subsection}

For our black box $B$, we use two dictionaries, both implemented as
balanced binary search trees: $B_1$ contains the at most $n$ distinct
elements seen so far, and $B_2$ stores \(O (\log m)\)-bit encodings of the
at most $n^\ell$ distinct $\ell$-tuples seen so far.  We search in each
dictionary once per query to $B$, that is, once per distinct \((\ell +
1)\)-tuple in $S$; we use a total of \(O (n^{\ell + 1} \log n) \subseteq O
(m)\) comparisons and time searching in $B_1$; we use a total of \(O
(n^{\ell + 1} \ell \log n) \subseteq O (\ell m)\) time searching in $B_2$,
but no comparisons between elements of $S$.

We maintain the invariant that, immediately before we process $s_i$, $B_1$
stores a set of pairs \(\langle a_1, r_1 \rangle, \ldots, \langle a_t, r_t
\rangle\), each of which consists of a distinct element \(a_j \in s_1,
\ldots, s_{i - 1}\) and $a_j$'s rank $r_j$.  We say $a_j$ has \emph{rank}
$r_j$ if it is the $r_j$th distinct element to appear in $S$; that is, for
some $k$, the first occurrence of $a_j$ is $s_k$ and there are $r_j$
distinct elements in \(s_1, \ldots, s_k\).  Notice operations on $B_1$ use
\(O (\log n)\) comparisons and time.  To query our black box $B$ before
processing $s_i$, we start by searching for $s_i$ in $B_1$.  If this
search succeeds, then we retrieve $s_i$'s rank; if it fails, then $s_i$ is
a new distinct element and we insert \(\langle s_i, r \rangle\), where $r$
is the number of distinct elements seen so far, including $s_i$.  After
this we have the ranks \(r_1, \ldots, r_{\ell - 1}\) of \(s_{i - \ell +
1}, \ldots, s_{i - 1}\), which $D_{s_{i - \ell}, \ldots, s_{i - 1}}$
stores, and the rank $r_\ell$ of $s_i$.

We also maintain the invariant that, immediately before we process $s_i$,
$B_2$ stores a set of pairs \(\langle g_1, p_1 \rangle, \ldots, \langle
g_{t'}, p_{t'} \rangle\), each of which consists of an \(O (\log m)\)-bit
encoding $g_j$ of a distinct $\ell$-tuple $\alpha$ in \(s_1, \ldots, s_{i
- 1}\) and a pointer $p_j$ to $D_\alpha$.  We use the \emph{gamma
code}~\cite{Eli75} to encode each \(a \in \alpha\), which encodes any
positive integer $x$ as a binary string \(\gamma (x)\) consisting of
$\lfloor \log x \rfloor$ copies of 0 followed by the \((\lfloor \log x
\rfloor + 1)\)-bit binary representation of $x$.

\begin{lemma}
\label{gamma_lemma}
We can encode any sequence \(x_1, \ldots, x_\ell\) of positive \(O (\log 
n)\)-bit integers as a unique \(O (\log m)\)-bit binary string.
\end{lemma}

\begin{proof}
The gamma code is prefix-free and, hence, unambiguous: any binary string
is the concatenation of at most one sequence of encoded integers.  Thus,
the encoding \(\gamma (x_1) \cdots \gamma (x_\ell)\) is unique and has 
length \(O (\ell \log n) \subseteq O (\log m)\).
\qed
\end{proof}

Notice operations on $B_2$ use \(O (\ell \log n)\) time and comparisons
between encodings of $\ell$-tuples, but no comparisons between elements of
$S$.  To query our black box $B$, after searching for $s_i$ in $B_1$, we
search for \(\gamma (r_1) \cdots \gamma (r_\ell)\) in $B_2$.  If this
search succeeds, then we retrieve a pointer to $D_{s_{i - \ell + 1},
\ldots, s_i}$; if it fails, then \(s_{i - \ell + 1}, \ldots, s_i\) is a
new distinct $\ell$-tuple, so we create a new statistics data structure
$D_{s_{i - \ell + 1}, \ldots, s_i}$ and insert \(\langle \gamma (r_1)
\cdots \gamma (r_\ell), p \rangle\) into $B_2$, where $p$ is a pointer to
$D_{s_{i - \ell + 1}, \ldots, s_i}$.

\bibliographystyle{plain}
\bibliography{sorting}

\begin{thebibliography}{10}

\bibitem{And91}
A.~Andersson.
\newblock A note on searching in a binary search tree.
\newblock {\em Software: Practice \& Experience}, 10:1125--1128, 1991.

\bibitem{BW94}
M.~Burrows and D.J. Wheeler.
\newblock A locally adaptive data compression scheme.
\newblock Technical Report 124, Digital Equipment Corporation, Palo Alto,
  California, 1994.

\bibitem{CLRS01}
T.~H. Cormen, C.~E. Leiserson, R.~L. Rivest, and C.~Stein.
\newblock {\em Introduction to Algorithms}.
\newblock MIT Press/McGraw-Hill, 2nd edition, 2001.

\bibitem{Eli75}
P.~Elias.
\newblock Universal codeword sets and representations of the integers.
\newblock {\em {IEEE} Transactions on Information Theory}, 21:194--203, 1975.

\bibitem{Gag04}
T.~Gagie.
\newblock Dynamic {Shannon} coding.
\newblock In {\em Proceedings of the 12th European Symposium on Algorithms},
  pages 359--370, 2004.

\bibitem{Knu98}
D.~E. Knuth.
\newblock {\em The Art of Computer Programming}, volume~3.
\newblock Addison-Wesley, 2nd edition, 1998.

\bibitem{Man01}
G.~Manzini.
\newblock An analysis of the {Burrows}-{Wheeler} transform.
\newblock {\em Journal of the {ACM}}, 48:407--430, 2001.

\bibitem{Meh77}
K.~Mehlhorn.
\newblock A best possible bound for the weighted path length of binary search
  trees.
\newblock {\em {SIAM} Journal on Computing}, 6:235--239, 1977.

\bibitem{MS76}
J.I. Munro and P.M. Spira.
\newblock Searching and sorting in multisets.
\newblock {\em {SIAM} Journal on Computing}, 5:1--8, 1976.

\bibitem{PM95}
O.~Petersson and A.~Moffat.
\newblock A framework for adaptive sorting.
\newblock {\em Discrete Applied Mathematics}, 59:153--179, 1995.

\bibitem{PD04}
M.~P\v{a}tra\c{s}cu and E.~Demaine.
\newblock Tight bounds for the partial-sums problem.
\newblock In {\em Proceedings of the 15th Symposium on Discrete Algorithms},
  pages 20--29, 2004.

\bibitem{SB02}
R.~Sedgewick and J.~Bentley.
\newblock Quicksort is optimal.
\newblock Presented at \emph{Knuthfest}, 2002.

\bibitem{Sha48}
C.~E. Shannon.
\newblock A mathematical theory of communication.
\newblock {\em Bell System Technical Journal}, 27:379--423, 623--656, 1948.

\bibitem{ST85}
D.~D. Sleator and R.~E. Tarjan.
\newblock Self-adjusting binary search trees.
\newblock {\em Journal of the {ACM}}, 32:652--686, 1985.

\end{thebibliography}

\end{document}